\documentclass[12pt]{article}
\usepackage[latin1]{inputenc}
\usepackage{slashed}
\usepackage{amsmath}
\usepackage{textcomp}
\usepackage{amssymb}
\usepackage{amsfonts}
\usepackage{indentfirst}
\usepackage{color}
\usepackage[top=2cm,bottom=2cm,left=3cm,right=2cm]{geometry}
\newcommand{\nin}{\noindent}
\newcommand{\be}{\begin{equation}}
\newcommand{\ee}{\end{equation}}
\newcommand{\bea}{\begin{eqnarray}}
\newcommand{\eea}{\end{eqnarray}}

\newcommand{\nn}{\nonumber\\}

\begin{document}
 
\nin KCL-PH-TH/2015-11

\begin{center} 
 
 {\Large{\bf Effective matter dispersion relation in quantum\\ covariant Ho\v{r}ava-Lifshitz gravity}}

\vspace{0.5cm}

{\bf J. Alexandre} and {\bf J. Brister}

\vspace{0.2cm}

King's College London, Department of Physics, WC2R 2LS, UK\\
{\small jean.alexandre@kcl.ac.uk, james.brister@kcl.ac.uk}

\vspace{1cm}

{\bf Abstract}

\end{center}
 
We study how quantum fluctuations of the metric in covariant Ho\v{r}ava-Lifshitz gravity influence the propagation of 
classical fields (complex scalar and photon).
The effective Lorentz-symmetry violation induced by the breaking of 4-dimensional diffeomorphism is then evaluated, 
by comparing the dressed
dispersion relations for both external fields. The constraint of vanishing 3-dimensional Ricci scalar
is imposed in the path integral, which therefore explicitly depends on two propagating gravitational degrees of freedom only. 
Because the matter fields are classical, the present model contains only logarithmic divergences.
Furthermore, it imposes the characteristic Ho\v{r}ava-Lifshitz scale
to be smaller than $10^{10}$ GeV, 
if one wishes not to violate the current bounds on Lorentz symmetry violation.

\vspace{2cm}

\section{Introduction}

Among the different approaches to Modified Gravity, Ho\v{r}ava-Lifshitz gravity (HL) consists in choosing different scaling properties of 
time and space coordinates, and therefore breaking 4-dimensional diffeomorphism \cite{HLinitial}. 
This feature allows to increase the number of space derivatives of the metric, 
while keeping the number of time derivatives minimal. This leads to a better ultraviolet (UV) behaviour of graphs, 
without introducing ghosts, and therefore provides a power-counting renormalisable theory of gravity.

A problem of the original model of HL gravity is the existence of an additional scalar degree of freedom (dof) for the metric, which
can be understood as a Goldstone mode arising from breaking of 4-dimensional diffeomorphism \cite{strongcoupling}.
A solution to this problem has been proposed in \cite{covariantHL}, where an auxiliary field $A$ is introduced, such that its ``equation of motion''
leads to an additional constraint, eliminating the unwanted scalar degree of freedom of the metric. 
The resulting theory is invariant under a new Abelian gauge symmetry $U(1)$ 
which involves the metric components, the auxiliary field $A$ and an additional auxiliary field. This gauge symmetry, together with 
the 3-dimensional diffeomorphism of HL gravity, can be shown to be equivalent to a 4-dimensional diffeomorphism at the lowest order in a 
post-Newtonian expansion, showing the equivalence with GR at long distances. For this reason this extension is called covariant HL gravity.  
We note however, that the long-distance limit is not obviously recovered: it has been shown in \cite{Abdalla} that the equivalence principle
is not automatically retrieved in the infrared, and that the meaning of the auxiliary
fields and their coupling to matter are still open questions.

Because of the anisotropy between space and time, HL gravity is naturally described in terms of the Arnowitt-Deser-Misner (ADM) 
decomposition of the metric, which expresses a space-time foliation. An important consequences of space-time anisotropy is the possibility
of imposing the lapse function $N$ to depend on time only. This situation is called the projectable case
and will be considered in this article.  
The non-projectable case has been studied in \cite{non-projectable} for static spherically symmetric solutions of covariant HL gravity.

This covariant extension to HL has led to several studies, including spherically symmetric solutions \cite{sphericalsymm} and their relevance to an 
alternative model for galaxy rotation curves \cite{MM}, cosmological 
solutions \cite{cosmo}, as well as theoretical and phenomenological consistency tests of the theory \cite{tests}.

\vspace{0.5cm}

In the present article, we study the effective dispersion relation for classical external fields (scalar and photon), coupled to the covariant 
extension of HL gravity, after
integrating over gravity degrees of freedom at one-loop, on a flat background. The constraint provided by the auxiliary 
field $A$ consists in 
setting the curvature tensor to zero, $R=0$, which we impose in the path integral over metric fluctuations. The restoring 
force for quantum fluctuations is
then provided by higher order space derivatives of the metric, and leads to a prediction of the Lorentz-symmetry  
violating effective dispersion relations for these fields. 

Studies of effective dispersion relations for Lifshitz-type models in flat space time have shown the limitations of 
phenomenological viability of these models. It was first noted in \cite{IengoRussoSerone}, where the unnatural fine tuning of bare parameters is shown, 
in order to  match the light cones seen by two different scalar particles interacting. Similar studies were done in \cite{ABH}, 
where the effective dispersion relation for interacting Lifshitz fermions is derived, in the case where flavour symmetry is broken. Similarly,
\cite{ABV} describes fermion effective dispersion relations in Lifshitz-type extensions of Quantum Electrodynamics.

The effective speed of light seen by matter interacting with HL gravity is studied in \cite{Pospelov}. The Authors derive this effective 
speed seen by a scalar field and an Abelian gauge field, and compare these to measure Lorentz-symmetry violation. 
We are doing a similar study here, treating matter as classical though, and imposing the constraint $R=0$ in the integration over graviton dof.
We note here that $R=0$ is not a gauge choice, but a constraint which has the physical effect to remove one degree of freedom in the theory.
As described below, the constraint leads 
to the vanishing of one of the scalar components of the space metric.

We note here related works, involving quantization of HL gravity. \cite{Saueressig1} uses Exact Renormalisation Group 
methods to provide an insight into the existence of asymptotically safe gravity \cite{safe}. 
Also, \cite{triangle} is based on causal dynamical triangulations to study phase transitions of space time geometry in (2+1)-dimensional HL gravity. 
Finally, \cite{quantumWang} describes how (1+1)-dimensional HL gravity can be quantised in a similar way as a harmonic oscillator. 
 
A short review on the covariant version of HL gravity is presented in the next section, and the coupling to matter fields is presented in section 3,
together with the integration of gravitons. This calculation is done for 
$\lambda\ne1$ (parameter in the kinetic term of gravitons), but the result does not depend on $\lambda$, which is a consequence of the vanishing 
of the trace of the fluctuating space metric. Section 4 presents a phenomenological analysis and our conclusions.

\section{Covariant Ho\v{r}ava-Lifshitz gravity}

\subsection{The action}

We consider the anisotropic scaling $z=3$, such that the mass dimensions of time and space are $[t]=-3$ and $[x]=-1$. In projectable HL gravity, 
the gravitational degrees of freedom are the lapse function $N(t)$, the shift function $N_{i}(x,t)$ and the 3-dimensional space metric $g_{ij}(x,t)$, 
which appear in the ADM form of the space-time metric
\begin{equation} \label{ADM}
ds^2=-c^2N^2dt^2+g_{ij} \left(dx^i+N^i dt\right)\left(dx^j+N^j dt\right)~,
\end{equation}
where $c$ is the speed of light, with dimension $[c]=2$. The mass dimensions of metric components are $[N]=0$, $[N_i]=2$, and $[g_{ij}]=0$.

The covariant version of HL gravity involves the auxiliary gauge fields $A(x,t)$ and $\nu(x,t)$ whose role is to impose a constraint which 
eliminates the scalar degree of freedom of HL gravity, as explained below. The action is then given 
by\footnote{It has been shown in \cite{daSilva} that the present extension to the original HL gravity is valid for any $\lambda$.} 
\be\label{action}
S=\frac{1}{\epsilon^2}\int dt d^3x\sqrt{g}\Big\lbrace N\left[ K_{ij}K^{ij}-\lambda K^2-V+
\nu\Theta^{ij}(2K_{ij}+\nabla_i\nabla_j\nu)\right]-AR\Big\rbrace ,
\ee
where 
\bea
K_{ij}&=&\frac{1}{2 N} \left\{\dot{g}_{ij}-\nabla_i N_j-\nabla_j N_i\right\},~~ i,j=1,2,3\\
\Theta^{ij}&=&R^{ij}-\frac{1}{2}R g^{ij}\nn
V&=&-c^2R-\alpha_1 R^2-\alpha_2 R^{ij}R_{ij}-\beta_1 R^3-\beta_2 R R^{ij} R_{ij}\nn
&&-\beta_3 R_i^j R_j^k R^i_k-\beta_4 R\nabla^{2}R-\beta_5 \nabla_i R_{jk} \nabla^i R^{jk}\nonumber~.
\eea
and the different tensors correspond to the 3-dimensional metric $g_{ij}$.
Note that $[A]=4$, $[\nu]=1$, $[\epsilon]=0$, and the potential $V$ includes all the renormalisable operators even under parity.
The dimensions of the various terms in the Lagrangian are
\be
[R]=2,~ [{R^2}]=4, ~[{R^3}]=[{\Delta R^2}]=6,~[c^2]=4,~[\alpha_i]=2,~[\beta_j]=0~.
\ee
Note that the term $ R^{ijkl}R_{ijkl}$ does not appear, as the Weyl tensor in three
dimensions automatically vanishes.\\

The action (\ref{action}) is invariant under the following transformations:
\begin{itemize}
\item 3-dimensional diffeomorphism 
\bea \label{diffeo}
\delta t&=&f(t)\\
\delta x^{i}&=&\xi^i(t,x)\nn
\delta g_{ij}&=&\partial_{i}\xi_j+\partial_{j}\xi_i+\xi^{k}\partial_{k}g_{ij}+f\dot{g}_{ij} \nn
\delta N_{i}&=&\partial_{i}\xi^kN_k+\xi^k\partial_kN_{i}+\dot{\xi^{j}}g_{ij}+\dot{f}N_i+f \dot{N}_{i} \nn
\delta N&=&\xi^k\partial_kN+\dot{f}N+f \dot{N} \nn
\delta A&=&\xi^k\partial_kA+\dot{f}A+f \dot{A} \nonumber
\eea
\item $U(1)$ symmetry
\bea
\delta_\alpha N&=&0\nonumber \\ 
\delta_\alpha g_{ij}&=&0 \nonumber\\
\delta_\alpha N_{i}&=&N \nabla_{i}\alpha \nonumber \\ 
\delta_\alpha A&=&\dot{\alpha}-N^{i}\nabla_{i}\alpha\nn
\delta_\alpha\nu&=&\alpha
\eea
where $\alpha$ is an arbitrary spacetime function.
\end{itemize}
In what follows, we shall need to distinguish tensors contracted with $g_{ij}$ from those contracted with the flat-space metric $\delta_{ij}$. 
For two vectors $v_i,w_i$, 
we will write the former as $v^iw_i$ and the latter as $v_iw_i$, which leads to $v^\mu w_\mu = -v_0 w_0 + v^i w_i$, where
\be
v^iw_i = v_iw_jg^{ij} = (g^{-1})^{ij}v_i w_j 
= \left(1- \epsilon h_{ij} + \epsilon^2 h_{ik}h_{jk}\right)v_i w_j + {\cal O}(\epsilon^3)~.
\ee
Also, we denote $v^2=v_iv_i$, we use $\partial_i$ for the flat 3-space derivative, and
$\partial^2$ for the flat 3-space Laplacian.

\subsection{Gauge fixing and $U(1)$-symmetry constraints}

An interesting way of counting the dof of the theory is given in \cite{Henneaux}, where the Authors use a Hamiltonian description of gauge theories.
They show that the number of primary constraints to take into account is the number of gauge functions {\it plus} 
the number of their time derivatives, in the situation where these gauge functions depend on space and time.
This is because gauge functions and their time derivatives must be considered independent, when defining a boundary condition for the 
evolution of gauge fields.
In our case, we have 10 independent metric components ($N,~N_i,~g_{ij}$) and we see from the gauge transformations (\ref{diffeo}) that the functions 
$\xi^i$ count twice since they appear with their time derivative, while $f$ counts once only because it depends on $t$ only.
The total number of dof is therefore $10-(2\times3+1) = 3$. 
We are then left with 3 dof, 
consistently with the appearance of the Goldstone boson after breaking of the
4-dimensional diffeomorphism to 3-dimensional diffeomorphism.

The metric fluctuations $h_{ij}$ are defined by $g_{ij} = \delta_{ij}+ \epsilon h_{ij}$ and,
choosing the synchronous gauge, where $N=1$ and $N_i=0$, we decompose $h_{ij}$ as
\be
h_{ij} =H_{ij} + \partial_i V_j +\partial_j V_i + (\frac{1}{3}\delta_{ij} - \frac{\partial_i \partial_j}{\Delta})B + \frac{1}{3}\delta_{ij}h,
\ee
where $H_{ij}$ is transverse traceless, $V_i$ is transverse and $h=$tr\{$h_{ij}\}$.

One can easily see that the variation of the action (\ref{action}) with respect to $A$ leads to $R=0$, 
which is a condition we will impose in the path integral
over graviton dof. This condition should be satisfied at the linear order in metric fluctuations, 
since we consider only quadratic terms in the gravity action.
We obtain then
\be
0=R=-\epsilon \frac{2}{3}\partial^2(B+ h) +O(\epsilon^2)~,
\ee
which, together with boundary conditions $h(\infty)= B(\infty)=0$ leads to $h=-B$ everywhere. We can re-write our expansion of $h_{ij}$ as
\be
h_{ij} =H_{ij} + \partial_i V_j +\partial_j V_i +  \frac{\partial_i \partial_j}{\Delta}h .
\ee

Finally, We fix the $U(1)$ symmetry by setting $\nu =0$, and we also note that ghosts decouple from matter at one loop, since the corresponding action is cubic in fluctuations of ghosts/gravitons.

\section{Effective dispersion relation for matter fields}

\subsection{Coupling to matter}

We now wish to couple the  gravity sector to classical matter, including a complex scalar field $\phi$ and an Abelian gauge field $A_\mu$.
In ~\cite{daSilva} a generic action is derived for the coupling of Covariant HL gravity to a scalar field, 
but we restrict ourselves to the case which recovers Lorentz symmetry in the IR
\bea
S_{scalar} = -\int d^3x dt~ \sqrt{g}( -\dot{\phi}\dot{\phi^\star}+c^2 g^{ij}\partial_i \phi \partial_j \phi^\star )~,
\eea
where $c$ is the speed of light, with mass dimension 2.\\
The coupling to an Abelian gauge field $A_\mu$ is described by the action
\be
S_{photon} = -\frac{1}{4}\int d^3x dt ~\sqrt{g} (-2 g^{ij}F_{0i}F_{0j} +c^2 g^{ik} g^{jl} F_{ij}F_{kl})~,
\ee
and we need not worry about gauge fixing for the Abelian field $A_\mu$, since it is considered an external source. 
We wish to calculate effective dispersion relations for the matter fields $\phi$ and $A_\mu$. As these fields couple to gravity 
only through their first derivatives, we may treat those derivatives as constant external fields:
\begin{itemize}
\item $\phi=\phi_0\exp(ik^\mu x_\mu)$ leads to $\partial_i\phi\partial_i\phi^\star=(\vec k)^2\phi_0^2$;
\item $A_i=A^0_i\sin(k^\mu x_\mu)$ leads to $F_{ij}^2=2(\vec k)^2(A^0)^2-2(\vec k\cdot\vec A^0)^2 + {\cal O}(k^4)$.
\end{itemize}
The (anisotropic) effective action will be given by an expression of the form
\be
S_{scalar}^{eff} = \int d^3x dt ((1+a)\dot\phi\dot\phi^\star - (1+b) c^2 \partial_i\phi \partial_i \phi^\star)~,
\ee
with a corresponding dispersion relation
\be
(1+a)k_0^2 = (1+b)c^2 k^2 ~,
\ee
where we assuming $a, b \ll 1$. If we note $v_\phi^2$ the product (phase velocity $\times$ group velocity), we have then
\be
\label{dispersion}
\frac{v_\phi^2}{c^2}-1=b-a + {\cal O}(\epsilon^3)
\ee
Similarly, for the photon we will obtain an effective action of the form
\be
S_{photon}^{eff} = -\frac{1}{4}\int d^3x dt(-2(1+a')F_{0i}F_{0i} +c^2 (1+b')F_{ij}F_{ij})~,
\ee
with the corresponding correction
\be
\frac{v_A^2}{c^2}-1=b'-a' + {\cal O}(\epsilon^3)~.
\ee
As noted in \cite{Pospelov}, the measurable quantity which violates Lorentz symmetry is the difference
\be
\delta v^2\equiv |v_A^2-v_\phi^2| = c^2|b' -b -a' +a|~.
\ee

\subsection{Features of the one-loop integration}

Taking into account the above gauge fixing conditions and the constraint $R=0$ in the path integral, 
the gravity sector expanded to second order in the metric fluctuations reads
\bea
S_{gravity} &=& -\frac{1}{4}\int d^3x dt~H_{ij}(x)\left(\partial_t^2 + 4 \alpha_2  \partial^4 -4\beta_5 \partial^6\right) H_{ij}  \nn
&&~~~~~~~~~~~~+ 2\partial_i V_j \partial_t^2 \partial_i V_j +(1- \lambda)h \partial_t^2 h +O(\epsilon)
\eea
Note that only the terms from  the potential with no powers of $R$ survive, and only those with two powers of $R_{ij}$ contribute at this 
level, as $R_{ij}$ is $O(\epsilon)$.
We can now see that we have only two propagating degrees of freedom (the two polarizations of $H_{ij}$), as expected.
We may treat the cases of photons and scalars separately at the order at which we are working, 
as they have no interactions other than through gravity.\\

For the scalar we have
\be
S_{scalar}=\int d^3x dt \left\{\mathcal{L}_{\phi0}  +
\epsilon^2 c^2 \bigg(- \frac{1}{3}    H_{ij}H_{ij} - \frac{2}{3} \partial_i V_j \partial_i V_j  -  \frac{1}{6} h^2  \bigg)\partial_k\phi \partial_k \phi^\star
 + {\cal O}(\epsilon^3)\right\}
\ee
where $\mathcal{L}_{\phi0}$ is proportional to $ -\dot{|\phi|}^2 + c^2\partial_i\phi \partial_i \phi^\star$, the usual relativistic Lagrangian density for the scalar.
Following \cite{Pospelov}, because of isotropy in space coordinates, terms of the form $T^{ij}\partial_i\phi\partial_j\phi^\star$ that are quadratic in the graviton fields have been replaced by 
$(1/3)T_i^i \partial_j\phi\partial^j\phi^\star$. Terms that mix different components of the graviton cannot contribute to our corrections at $O(\epsilon^2)$ and so are neglected here. 
Terms linear in the graviton fields do not contribute here either, as we treat the scalars as classical external fields; for the same reason, we are able to integrate the quadratic terms by parts.

Similarly, for the photon we have 
\be
S_{photon} =\frac{1}{4}\int d^3x dt \left\{\mathcal{L}_{A0} 
+ \epsilon^2 c^2 \bigg(- \frac{1}{6} H_{ij}H_{ij} - \frac{1}{3} \partial_i V_j\partial_i V_j  - \frac{1}{6} h^2 \bigg) F_{kl}F_{kl}  
 + {\cal O}(\epsilon^3)\right\} ~,
\ee
where isotropy has been used and $\mathcal{L}_{A0}$ is proportional to the relativistic Lagrangian for photons.

\subsection{Integration}

As the components of the graviton do not mix, we may consider their contributions separately. 

\subsubsection{Spin 2}

As we seek the \emph{difference} between the modifications to the space and time components, we can neglect in the action 
the term $\mathcal{L}_{\phi0}$. We therefore need only to consider the action 
\be
\tilde S_{scalar} = -\frac{1}{4}\int d^3x dt ~H_{ij}(x)\left(\partial_t^2 + 4 \alpha_2  \partial^4 - 4\beta_5 \partial^6 
+ \frac{4\epsilon^2}{3}c^2\partial_k\phi \partial_k \phi^\star \right)H_{ij}(x)~.
\ee
for the scalar and
\be
\tilde S_{photon} =-\frac{1}{4} \int d^3x dt ~H_{ij}(x)\left(\partial_t^2 + 
4 \alpha_2  \partial^4 - 4\beta_5 \partial^6 +\frac{2\epsilon^2}{3}c^2 \frac{1}{4} F_{kl}F_{kl} \right)H_{ij}(x)~.
\ee
for the photon. 
\\

\nin{\it Scalar field}\\
We denote by $D(q)$ the Fourier transform of 
$\partial_t^2 + 4 \alpha_2  \partial^4 - 4\beta_5 \partial^6$ , and we have
\be
S_{H}=-\frac{1}{4} \int \frac{d^3p dp_0}{(2\pi)^4}\frac{ d^3qdq_0}{(2\pi)^4} H_{ij}(p)\left( D(q)
+\frac{4\epsilon^2}{3}c^2\partial_k\phi \partial_k \phi^\star \right)\delta(p+q)~H_{ij}(q)~.
\ee
Integrating over the two components of $H$, this gives a contribution to the partition function of
\bea
\label{det}
&&\left(\mbox{Det}\left\{(D(q)+\frac{4\epsilon^2}{3}c^2\partial_k\phi \partial_k \phi^\star)\delta(p+q)\right\}\right)^{-1} \\
&=& \exp\left\{ - \mbox{Tr} \left[ \ln\left(D(q)\delta(p+q)\right) 
+ \frac{4\epsilon^2}{3} c^2\partial_k\phi \partial_k \phi^\star~D^{-1}(q)\delta(p+q) \right] \right\}\nn
&=&\exp\left\{- \frac{4\epsilon^2}{3}c^2\partial_k\phi \partial_k \phi^\star 
\delta(0)\int \frac{d^3p dp_0}{(2\pi)^4}  \frac{1}{-p_0^2 +4\alpha_2 p^4 + 4\beta_5 p^6} + \cdots\right\}\nonumber~,
\eea
where $\delta(0)$ is a constant global volume factor, and dots represent field-independent terms. The integral in the above leads to:
\bea
\label{integral}
I = -\delta(0) \frac{4\epsilon^2}{3}c^2\partial_k\phi \partial_k \phi^\star \int\frac{d^3p dp_0}{(2\pi)^4}
\frac{1}{-p_0^2 +4\alpha_2 p^4 + 4\beta_5 p^6} \nn
=  - \delta(0)\frac{4\epsilon^2}{3(2\pi)^4}c^2\partial_k\phi \partial_k \phi^\star \times 
\frac{-i\pi}{2} \int \frac{d^3p}{p^2 \sqrt{\alpha_2 + \beta_5 p^2}}~,
\eea
and is logarithmically divergent. We regularize this integral by dimensional regularization, as it respects the symmetries of the theory
\bea
I_{a} = - \delta(0)\frac{4\epsilon^2}{3(2\pi)^4}c^2\partial_k\phi \partial_k \phi^\star \times 
\frac{-i\pi}{2} \int \frac{(3-a)\pi^{\frac{3-a}{2}} p^{2-a} dp}{\Gamma(\frac{5-a}{2}) p^2 \sqrt{\alpha_2 + \beta_5 p^2}} \mu^a \nn
= i\delta(0)\frac{\epsilon^2}{6\pi^2}c^2\partial_k\phi \partial_k \phi^\star \frac{1}{\sqrt{\beta_5}} \frac{\mu^a}{a}
+ \mbox{finite} ~,
\eea
where $\mu$ is an arbitrary mass scale. \\
After dividing by $i\delta(0)$, in order to take into account the Wick rotation and the space time volume, the 
identification with the speed (\ref{dispersion}) gives
\be\label{res1}
\frac{v_\phi^2}{c^2} -1= \frac{-\epsilon^2}{6\pi^2\sqrt{\beta_5}} \frac{\mu^a}{a} ~.
\ee

\nin {\it Photon}\\
Comparing the coefficients of the relevant terms in the actions, one can see the effective change in velocity for the 
photon will be 1/2 times that of the scalar. This leads to
\be\label{res2}
\frac{v_A^2}{c^2}-1 =  \frac{-\epsilon^2}{12\pi^2\sqrt{\beta_5}} \frac{\mu^a}{a} ~.
\ee
Note that both these results are sub-luminal, as might intuitively be expected: small fluctuations of the spatial metric about 
flat space should generally lead to a ``longer" path between two points.

\subsubsection{ Spin 1}

The spin 1 terms are similar to the spin 2 terms shown above, with $2\partial_i V_i \partial_i V_i$ in the place of $H_{ij}H_{ij}$. However, the calculation leads to an integral of the form
\be
\int \frac{d^3p dp_0}{(2\pi)^4} \frac{1}{p_0^2}~.
\ee
Unlike the previous case, the $p$ integral here is the integral of a polynomial, which can be formally taken to vanish under dimensional regularization \cite{Leibbrandt}.
The vanishing or finiteness of a regularised integral which otherwise would naively be divergent is explained pedagogically in \cite{Weinzierl}: in the regularised integral, 
divergences associated to different regions of the domain of integration cancel each other, such that the integral is finite when the regulator is removed.

\subsubsection{ Spin 0}
The coefficients of the relevant terms and thus the modifications to the velocities coming from the spin 0 component of the graviton, $h$, are equal for the 
scalar and photon.\footnote{This is unsurprising, see the similar calculation for regular Ho\v{r}ava Lifshitz gravity in \cite{Pospelov}, 
the field the authors call $\sigma$ vanishes here.} Hence the final result below does not depend on the parameter $\lambda$, which only appears in the spin 0 kinetic term.\\

\section{Analysis and conclusions}

From the results (\ref{res1}) and (\ref{res2}), the total measurable difference in the squared velocities is 
\be
\delta v^2\equiv v_\phi^2-v_A^2 = \frac{-\epsilon^2 c^2}{12\pi^2 \sqrt{\beta_5}}   \frac{\mu^a}{a}+ \mbox{finite} ~.
\ee
We define the beta function, in the limit $a \rightarrow 0$, by
\be
\beta = \mu \frac{\partial (\delta v^2)}{\partial \mu} =  \frac{-\epsilon^2 c^2}{12\pi^2 \sqrt{\beta_5}}~,
\ee
and we can then write, for some mass scale $\mu_0$,
\be
\label{res}
\delta v^2 = \frac{-\epsilon^2 c^2}{12\pi^2 \sqrt{\beta_5}} \ln\frac{\mu}{\mu_0}~.
\ee
In order to chose $\mu_0$, we repeat the calculation of the integral (\ref{integral}), regularised by the high momentum cut-off $\Lambda$ to obtain
\bea
I_{\Lambda} = i\delta(0) \frac{\epsilon^2}{6\pi^2}c^2\partial_k\phi \partial_k \phi^\star \frac{1}{\sqrt{\beta_5}} 
\ln \left(\sqrt\frac{\beta_5}{\alpha_2}\Lambda + \sqrt{1+\frac{\beta_5}{\alpha_2} \Lambda^2}\right) \nn
= i\delta(0)\frac{\epsilon^2}{12\pi^2}c^2\partial_k\phi \partial_k \phi^\star \frac{1}{\sqrt{\beta_5}}  \ln \frac{\Lambda^2}{\alpha_2} 
+ \mbox{finite} ~,
\eea
such that the identification with the speed (\ref{dispersion}) gives
\be
\frac{v_\phi^2}{c^2} -1= \frac{-\epsilon^2}{12\pi^2\sqrt{\beta_5}} \ln\frac{\Lambda^2}{\alpha_2} ~,
\ee 
and suggests $\mu_0=\sqrt{\alpha_2}$ as a natural choice.

The result (\ref{res}) is obtained in anisotropic Minkowski space time, and we rescale the time coordinate as 
$t\to t~M_{HL}^2$,
where $M_{HL}$ is a scale characteristic of HL gravity, below which the classical model can be considered relativistic. 
Speeds are then rescaled
as $v\to v~ M_{HL}^{-2}$, and we set the speed of light in isotropic Minkowski space time to 1 (we therefore identify the 
dimensionful quantity $c= M_{HL}^2$). 
The coupling constant appearing in the action (\ref{action}) is 
then $\epsilon=M_{HL}/M_{Pl}$, where $M_{Pl}$ is the Planck mass, and the measurable deviation from Lorentz symmetry in 
the effective theory is
\be\label{final}
\delta v^2=|v_\phi^2-v_A^2| =\frac{M^2_{HL}}{24\pi^2M^2_{Pl} \sqrt{\beta_5}}  \ln\frac{\mu^2}{\alpha_2}~,
\ee
which should be less than about $10^{-20}$ \cite{bounds}.

In order to get an idea of 
the order of magnitude for $M_{HL}$, one can set the different parameters to natural values in the present context, which 
are $\beta_5\simeq1$, $\alpha_2\simeq M_{HL}^2$ and $\mu\simeq M_{Pl}$. This shows that
one should respect $M_{HL}\lesssim10^{10}$ GeV for the result (\ref{final}) to satisfy the upper bounds on Lorentz-symmetry violation.
This value is also obtained in \cite{AL3},
where non-relativistic corrections to matter kinetic terms are calculated in the framework of another 4-dimensional diffeomorphism
breaking gravity model, and where $10^{10}$ GeV corresponds to the cut off above which the model is no longer valid.
We note that this scale also corresponds to the Higgs potential instability \cite{Higgsinst}, which could be
avoided by taking into account curvature effects in the calculation of the Higgs potential \cite{rajantie}, and it would be interesting
to look for a stabilizing mechanism in the framework of non-relativistic gravity models.

We comment here on the relevance of Lorentz-symmetry violation in the study of cosmic rays, 
with energies necessarily lower than the Greisen-Zatsepin-Kuzmin cutoff $E_{GZK}=10^{19.61 \pm 0.03}$eV 
\cite{Abraham:2010mj}, since the latter is of the order of the bound we find for $M_{HL}$. 
Above this energy, protons interact with the Cosmic Microwave Background, producing pions which decay and generate showers 
observable on Earth. 
As noted in \cite{sibiryakov}, resulting photon-induced showers would be highly sensitive to Lorentz-symmetry violation, 
and the observation of $10^{19}$eV photons would put strong bounds on the different Lorentz-symmetry violating parameters 
of the Standard Model 
Extension - SME -\cite{SME}. The observation of such photons would also help put bounds on  
parameters in the expression (\ref{final}), which could then be related to the SME.
The bound $\delta v^2\lesssim10^{-25}$ found in \cite{sibiryakov} from potential ultra-high-energy photons, is actually 
much smaller than 
the one considered above, and concerns electron/positrons created by these high energy photons in the presence of the Earth 
magnetic field. 
Taking into account this bound, and assuming $\beta_5$ of order 1,
we find the typical upper bound $M_{HL}\lesssim10^6$ GeV. Although this is still well above the current accessible energies at CERN, 
the observation of $\sim10^{19}$ eV photons
would definitely put much stronger constraints on Lorentz-symmetry violation.

We also remark that supersymmetric models have been studied in the context of Lifshitz-type theories \cite{HLsusy}, 
where an interesting feature is that particles in the same supermultiplet see the same limiting speed. Furthermore, in the case 
of supersymmetric gauge theory, the limiting speed is the same for matter and gauge supermultiplets. From the phenomenological point of view, 
it is suggested that the Lorentz-symmetry violation scale $M_{HL}$ should be above the SUSY breaking scale, in order to avoid fine-tuning problems imposed 
by bounds on Lorentz-symmetry violation. This means that $M_{HL}$ should be at least of the order of 10 TeV, 
which is well below the bound we find here.

We make a final comment regarding the calculation carried out in \cite{Pospelov}, for the original Ho\v{r}ava-Lifshitz gravity and 
where a \emph{quadratic} divergence is found. 
This stems not from the extra degree of freedom, but from treating the matter as quantum fields and then 
considering terms quartic in the matter fields, 
obtained from completing the square for the coupling terms linear in the graviton fields, that we neglected above. 
In terms of Feynman diagrams, treating 
matter as quantum fields consists in considering self energy graphs with an internal matter propagator. Our analysis 
shows that, as long as matter is classical, 
only logarithmic divergences arise.
Studies similar to the present one are planned, involving higher-order space derivatives of matter fields, in order 
to generate dynamically Lifshitz-type theories in flat space time,
as the ones analysed in \cite{ABH,ABV} for example.

\end{document}